## PAPER

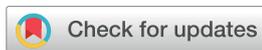

Check for updates

Cite this: DOI: 10.1039/d5sm00057b

# Numerical insights on the volume phase transition of thermoresponsive hollow microgels

Leah Rank  [ab] and Emanuela Zaccarelli  *[ab]

Hollow microgels, consisting of a pNIPAM polymer network with a central cavity, have significant potential due to their tunable softness and encapsulation capabilities. Using molecular dynamics simulations, we thoroughly characterise the swelling behaviour of neutral hollow microgels across the volume phase transition (VPT) upon varying crosslinker concentration, shell thickness, and size. In particular, we examine in detail the onset of cavity filling and its relation to the VPT, detecting the presence of a discontinuity in the radius of gyration of the microgels, if an appropriate balance between shell stiffness and thermoresposiveness is reached. The discontinuity is, however, absent in the behaviour of the hydrodynamic radius, in agreement with experimental observations. We then test our numerical model by direct comparison of form factors with available measurements in the literature and also establish a minimal-size, stable hollow microgel for future computationally feasible bulk investigations. Overall, our findings provide valuable insights into the fundamental swelling properties of hollow microgels that can be useful to control the opening and closing of the cavity for application purposes.

## 1 Introduction

Microgels are synthetic polymer networks,[1] typically ranging from a few tenths of nanometers up to a few microns in size. Depending on the constituent polymers, they display reversible swelling properties in response to environmental changes, *e.g.*, temperature,[2–4] pH[5,6] or salt concentration.[7,8] In particular, thermoresponsive microgels undergo a so-called volume phase transition (VPT), shifting from a swollen to a collapsed state at a specific temperature. This transition reflects the underlying coil-to-globule transformation of the polymer chains, induced by worsening solvent conditions as the temperature increases. Microgels made of poly-*N*-isosopropylacrylamide (pNIPAM) are among the most studied ones, with the VPT occurring at around 32 °C.[9] The combination of softness and responsiveness makes them ideal systems for fundamental condensed matter research[10–12] and for numerous applications in material science that exploit their controllable bulk behaviour.[13–16]

Among the different topologies, hollow microgels, *i.e.*, polymer network shells with an internal cavity, stand out as particularly intriguing. Even more than their non-hollow counterparts, these structures hold great potential as carriers encapsulating smaller nanoparticles, such as drugs, that require controlled release to specific areas in the body.[17–19] Furthermore, hollow microgels are of fundamental interest for their similarities to vesicles and cells,[20] due to their unique elastic properties allowing them to mimic the mechanical behaviour of these biological systems. Notably, the buckling of hollow microgels has been observed under specific osmotic conditions.[21] This phenomenon involves a shape transformation where the initially spherical microgel structure collapses asymmetrically, forming a double-layered, bowl-like morphology. Driven by osmotic pressure gradients, this transition results in a concave configuration with a distinctive inward folding that imparts a characteristic double-walled structure to the microgel.[19,21] This is analogous to what has been seen in soft vesicles[22] and red blood cells,[23] where soft particles deform under certain mechanical stresses.

However, to ensure efficient applications, it is important to tune the cavity size relative to the shell thickness and to understand how this ratio influences the behaviour as a function of temperature. In this context, Dubbert *et al.*[17] explored different experimental conditions to monitor the cavity's stability, even above the VPT temperature (VPTT). They found that preserving and stabilising the hollow structure comes at the expense of efficient thermoresponsive behaviour. Another control parameter for the rigidity is the crosslinker concentration within the shell, which has been studied by the formerly mentioned work as well as Contreras-Cáceres *et al.*[24] who used various microscopy techniques to identify the microgel's shape in detail. As expected, these authors found that increasing the crosslinker concentration helps to maintain the cavity open at high temperatures, causing the microgel to be less responsive to temperature increase. The aim to further examine hollow microgels and their swelling behaviour is therefore to be able to tune their stiffness such that they maintain their hollowness up to the desired temperatures, but still respond to temperature variation in terms of shrinking. For some specific purposes, it may also be useful to tune the occurrence of the filling of the cavity below or above the VPT temperature. Most of the available experimental studies of hollow microgels are based on the use of a degradable core on which a poly(N-isopropylacrylamide) (pNIPAM) shell is built.[17,19,24–26] This method allows for producing almost monodisperse hollow microgels. Other studies have employed an alternative protocol, where the microgel shell is synthesised around an inner microgel, rather than a solid core, which is also subsequently dissolved.[27,28] Comparisons of the hollow microgels obtained this way with numerical simulations[28] indicate that the resulting shell is significantly less dense in monomers than standard microgels, likely due to the complex interplay between the two networks. In contrast, an extensive numerical characterisation of hollow microgels synthesised around a solid core, particularly regarding their behaviour across the VPT, and a detailed comparison to experiments, are still lacking.

[a] CNR Institute of Complex Systems, Uos Sapienza, Piazzale Aldo Moro 2, 00185 Roma, Italy. E-mail: emanuela.zaccarelli@cnr.it
[b] Department of Physics, Sapienza University of Rome, Piazzale Aldo Moro 2, 00185 Roma, Italy

The present study aims to fill this gap by examining the single-particle properties of neutral hollow microgels with varying shell thicknesses, total sizes, and crosslinker concentrations to be able to obtain tunable hollowness with added thermoresponsive character using the right balance of the synthesis parameters. We test our numerical model by comparing the numerical form factors with available experimental ones for both very large microgels[19] and for more standard ones,[17] measured respectively by static light scattering and by small angle neutron scattering, for two different temperatures above and below the VPT. Additionally, we also assess size effects to stabilise the inner cavity of the hollow microgels while using a minimal number of monomers, paving the way for future computationally feasible investigations of their collective behaviour. The main findings of our study are devoted to unveiling the complex interplay between the occurrence of the VPT, influenced by solvent–polymer interactions, and the closing of the cavity, which needs to overcome the elastic response of the network, as a function of the various synthesis parameters mentioned above. We present evidence of specific conditions—namely, a high crosslinker concentration combined with a moderate shell thickness—under which the size of the microgel, expressed by the radius of gyration, exhibits a peculiar discontinuity as a function of temperature, signaling the filling of the cavity. Varying the combinations of parameters, we find situations where the hole disappears even before the VPTT, or it does not close at all up to extremely elevated temperatures, well beyond those studied in the experiments. Our results are rationalized in the context of existing experimental findings,[17,19,24] thereby offering a useful map of the different parameter combinations that can be employed to achieve the preservation of the hollowness up to the desired temperatures.

# 2 Model and methods

## 2.1 Assembly

Our numerical protocol to assemble disordered polymer networks is based on previous works,[29] exploiting the oxDNA simulation package.[30] Specifically, we use a binary mixture of patchy particles, representing monomers as bivalent particles and crosslinkers as tetravalent ones. All the particles have mass $m_m$ and diameter $\sigma_m$, which serve as the units of mass and length, respectively. During the molecular dynamics (MD) simulations of the assembling process, all particles are confined within a sphere using a stiff harmonic potential acting at radius $Z_{out}$. The molar concentration of the crosslinkers $c$ is varied between 1% and 10% according to most common experimental values.

To make the microgels hollow, all particles are also prevented from entering an inner spherical region[28] via a gravitational force acting symmetrical around the center of the outside barrier, with radius $Z_{in} < Z_{out}$, as illustrated in Fig. 1.

After almost all of the bonds ($\gtrsim 99.8\%$) are formed, most of the particles are found in one big cluster, and the remaining particles not belonging to the big network (consisting of $N_m = \rho_m \cdot (4\pi/3)(Z_{out}^3 - Z_{in}^3)$ monomers) are removed.

The total number of particles $N_m$ is chosen to achieve the number density $\rho_m \approx 0.08$, as established by successful comparison to experiments for standard microgels.[19,31] We note that in ref. 28, the number density used for hollow microgels synthesised around an inner microgel is approximately $\rho_m \approx 0.03$, significantly lower than in our current study due to interpenetration with the inner network. Here, the solid inner core does not extend into the shell region, therefore, we maintain the density consistent with standard microgels. While standard microgels require an additional force on the more reactive crosslinkers to sustain the core-corona structure, we assume that this additional force can be neglected here, given the reduced assembly volume of the shell.

## 2.2 Microgel simulations

Independently of the employed assembly method, once the network is formed, the attractive patches are replaced by permanent bonds,[29] represented by the well-known bead-spring model of Kremer and Grest.[32] In particular, non-bonded monomers experience an excluded volume interaction, modeled as a Weeks–Chandler–Andersen (WCA) potential,

$$V_{\text{WCA}}(r) = \begin{cases} 4\varepsilon\left(\left(\dfrac{\sigma_{\text{m}}}{r}\right)^{12} - \left(\dfrac{\sigma_{\text{m}}}{r}\right)^6\right) + \varepsilon & \text{if } r \leq 2^{1/6}\sigma_{\text{m}} \\ 0 & \text{otherwise,} \end{cases} \quad (1)$$

while the bonded monomers additionally interact with the finite extensive nonlinear elastic (FENE) potential allowing for unbreakable bonds,

$$V_{\text{FENE}}(r) = -\varepsilon k_{\text{F}} R_0^{\,2} \ln\left(1 - \left(\dfrac{r}{R_0\sigma_{\text{m}}}\right)^2\right), \quad \text{if } r < R_0\sigma_{\text{m}}. \quad (2)$$

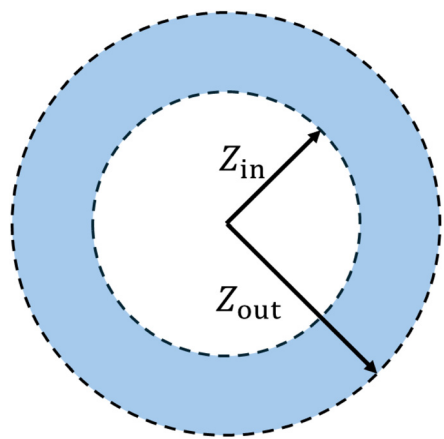

**Fig. 1** Sketch of the borders during the assembly process. The patchy particles are confined within a shell of outer radius $Z_{out}$ and inner radius $Z_{in}$.

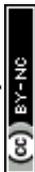

The parameter $\varepsilon$ defines our unit of energy, $k_{\text{F}} = 15$ is the spring constant, and $R_0 = 1.5$ is the maximum bond extension. To model the effect of increasing temperature, we use a solvophobic potential[33,34] that models the increasing repulsion to the solvent as an effective attraction between the monomers:

$$V_\alpha(r) = \begin{cases} -\varepsilon\alpha & \text{if } r \leq 2^{1/6}\sigma_{\text{m}} \\ \dfrac{1}{2}\varepsilon\alpha\left[\cos(\gamma(r/\sigma)^2 + \beta) - 1\right] & \text{if } 2^{1/6}\sigma_{\text{m}} < r \leq R_0\sigma_{\text{m}} \\ 0 & \text{otherwise} \end{cases} \quad (3)$$

with $\beta = 2\pi - 2.25\gamma$, $\gamma = \pi/(2.25 - 2^{1/3})$. The parameter $\alpha$ acts as an effective temperature: for $\alpha = 0$, the microgel is in good solvent conditions, while the volume phase transition (VPT) occurs around $\alpha \approx 0.65$.[35,36] MD simulations are conducted at a fixed temperature, $T^* = k_{\text{B}}T = 1.0$, where $k_{\text{B}}$ is the Boltzmann constant. To ensure a constant temperature, the Nosé–Hoover thermostat[37] is used, following a leapfrog integration scheme with a time step $\delta t^* = \delta t\sqrt{\varepsilon/(m_{\text{m}}\sigma_{\text{m}})} = 0.002$. The simulations have been conducted for each microgel individually, lasting up to $10^7$ time steps.

We characterise the hollow microgels of various sizes using a relative shell thickness, defined as $\delta_{\text{rel}} = (Z_{\text{out}} - Z_{\text{in}})/Z_{\text{out}}$. Further, we analyse the differences between the structure directly after assembly, where all particles lie between $Z_{\text{in}}$ and $Z_{\text{out}}$, and the relaxed structure after equilibration, *i.e.*, evolution under the Kremer–Grest potential without boundaries, to determine how the microgel structure evolves depending on the initial assembly parameters.

### 2.3 Calculated observables

The swelling curve of a microgel is monitored by calculating its radius of gyration from equilibrated configurations, defined as

$$R_{\text{g}} = \left\langle \sqrt{\dfrac{1}{N_{\text{m}}}\sum_i^{N_{\text{m}}}(\vec{r}_i - \vec{r}_{\text{cm}})^2} \right\rangle, \quad (4)$$

where $\vec{r}_i$ refers to the position of the $i$-th monomer and $\vec{r}_{\text{cm}}$ to the microgel center of mass.[38] The angled brackets denote an average over several configurations at different uncorrelated times of the equilibrated network. Alternatively, and more in line with experimental practices, the microgel size at different temperatures can be described by the evolution of the hydrodynamic radius $R_{\text{H}}$ with temperature. Several definitions have been proposed in the recent past, but we calculate $R_{\text{H}}$ as:

$$R_{\text{H}} = \left\langle 2\left[\int_0^\infty \dfrac{1}{\sqrt{(a^2 + \theta)(b^2 + \theta)(c^2 + \theta)}}\mathrm{d}\theta\right]^{-1} \right\rangle \quad (5)$$

where $a, b, c$ are the principal semi-axes of the gyration tensor of the convex hull enclosing the microgel, which is assumed to be instantaneously ellipsoidal in shape. This relation was put forward by Hubbard *et al.*[39] to calculate the hydrodynamic radius of differently shaped Brownian particles and was recently validated against experiments of non-hollow microgels in ref. 40.

We also evaluate the radial density distribution $\rho(r)$ as a function of the distance $r = |\vec{r}|$ from the center of mass of the microgel, given by

$$\rho(r) = \left\langle \sum_{i=1}^{N_{\text{m}}} \delta(|\vec{r} - \vec{r}_i|) \right\rangle. \quad (6)$$

Finally, we calculate its representation in Fourier space, namely the form factor:

$$P(q) = \dfrac{1}{N_{\text{m}}}\left\langle \sum_{i,j} \exp(-i\vec{q}\cdot\vec{r}_{ij}) \right\rangle, \quad (7)$$

which is essential for comparison with experiments. The vector $\vec{r}_{ij} = \vec{r}_i - \vec{r}_j$ denotes the distance between monomer $i$ and $j$ within the microgel, and the corresponding sum runs over all particle pairs.

### 2.4 Form factor fits

To extract the density profiles from experimental data, it is necessary to perform adequate fits of the form factors. We can thus exploit simulations, where both quantities can be calculated directly, to evaluate the appropriateness of different fitting functions. Therefore, we first analyse the density profiles to identify a suitable functional form, which can then be converted *via* Fourier transformation to obtain a fitting function for the corresponding form factor. In order to provide a recipe to be followed also in experiments, we privilege the use of analytic functions for the density profiles containing significant parameters that sufficiently describe the microgel's topology.

More specifically, we find that a suitable representation for the density profile of all simulated microgels is given by a

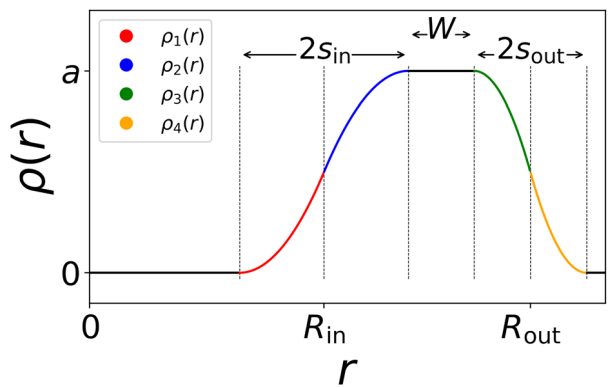

**Fig. 2** Sketch of the density profile made up of four different consecutive parabolas, inspired by ref. 41. From 0 to $R_{in} - s_{in}$, $\rho(r)$ is zero indicating the cavity region. The shell, roughly found between $R_{in}$ and $R_{out}$, shows fading edges inward and outward analogously to the non-hollow microgel with a homogeneous core and a fuzzy corona.

parabolic description,[41] which reads as

$$\rho_{parab}(r) = \begin{cases} \dfrac{a}{2} \dfrac{(r - (R_{in} - s_{in}))^2}{s_{in}^2} & \text{if } R_{in} - s_{in} < r \leq R_{in} \\[2mm] a - \dfrac{a}{2} \dfrac{(r - (R_{in} + s_{in}))^2}{s_{in}^2} & \text{if } R_{in} < r \leq R_{in} + s_{in} \\[2mm] a - \dfrac{a}{2} \dfrac{(r - (R_{out} - s_{out}))^2}{s_{out}^2} & \text{if } R_{out} - s_{out} < r \leq R_{out} \\[2mm] \dfrac{a}{2} \dfrac{(r - (R_{out} + s_{out}))^2}{s_{out}^2} & \text{if } R_{out} < r \leq R_{out} + s_{out} \\[2mm] 0 & \text{otherwise.} \end{cases}$$

(8)

This equation consists of four consecutive parabolas ($\rho_i(r)$ with $i = 1,2,3,4$), schematically plotted in Fig. 2 which visually explains the different defining parameters $a$, $s_{in}$, $s_{out}$, $R_{out}$, and $R_{in} = R_{out} - s_{out} - W$.

To obtain the form factor, we first need to calculate the amplitude

$$A(q) = \int \rho_{fit}(r) e^{-i\vec{q}\cdot\vec{r}} d\vec{r}$$

(9)

and hence each parabola can be analytically Fourier transformed to eventually get $A_{parab}(q) = A_{parab}(q, R_{out}, s_{in}, s_{out}, a)$, as detailed in the Appendix, giving a fitting function for $P(q)$:

$$P_{parab}(q) = \frac{A_{parab}(q)^2}{N_m^2} + \frac{I_0}{1 + \xi^2 q^2}$$

(10)

where a Lorentzian term with two extra fit parameters, the amplitude $I_0$ and the polymer network's average correlation length $\xi$, is added, as in experiments, to take into account the finite size of the microgel.[42] The normalisation is given by $N_m^2$ which is exactly known in simulations. The form factor can thus be fitted using all the fit parameters mentioned above that define the parabolic density profile as well as the two terms from the Lorentzian. Alternatively, we can simply take the values from the fit parameters obtained from $\rho_{parab}$ and fit only the Lorentzian part. As shown in the following, the results of the two approaches are virtually identical.

In addition, we notice that when we simulate small hollow microgels, the density profiles can be captured in two ways: either *via* the parabolic approach described above with $W = 0$ or due to the almost fully symmetric shape even *via* a Gaussian density profile of width $\sigma$:

$$\rho_{Gauss}(r) = C \cdot \exp\left(-\frac{(r - \mu)^2}{2\sigma^2}\right).$$

(11)

Since this is a symmetric function, it requires even fewer fit parameters. Here, the form factor can be calculated with the formula put forward by Gradzielski *et al.*,[43] reading as,

$$A_{Gauss}(q) = C \cdot \frac{4\pi}{q} \exp\left(-\frac{q^2\sigma^2}{2}\right) \left[\mu \sin(q\mu) + q\sigma^2 \cos(q\mu)\right]$$

(12)

Next, the amplitude $A_{Gauss}(q)$ can be used in eqn (10) the same way as for the parabolic case. To be more precise, for the smaller microgel in its hollow state, $W = 0$ is fixed, thus we have one less fit parameter, while $R_{in} = R_{out} - s_{out} - W = R_{in} = R_{out} - s_{out} - s_{in}$ can be calculated the same way as before, and the Gaussian mean value is approximately equivalent to $\mu = (R_{out} + R_{in})/2$, respectively $\sigma = (s_{out} + s_{in})/2$.

At high temperatures, when the microgels experience a collapse and the cavity vanishes, the above-mentioned fits no longer hold. In that case, we resort to the standard treatment of microgel structures in terms of the fuzzy sphere model[3] which has been used in previous works[29] to describe density profiles of non-hollow microgels.

## 3 Results

### 3.1 Swelling behaviour of hollow microgels: a typical signature in the radius of gyration

Hollow microgels exhibit a similar deswelling behaviour upon temperature increase as their non-hollow counterparts, driven by the decreased solvating ability of the background liquid. However, due to their internal cavity, hollow microgels display an additional structural feature: the inner cavity collapses at elevated temperatures and low enough shell elasticity. We begin by demonstrating this unique swelling and collapse behaviour in a specific hollow microgel. For this purpose, we select the representative case where $\delta_{rel} = 0.275$ and $c = 5\%$. Snapshots at different temperatures are shown in Fig. 3(a), and the corresponding radial density distributions $\rho(r)$ are plotted in Fig. 3(b).

It is evident that up to $\alpha = 0.6$, just below the VPT, the microgel maintains its cavity intact, while eventually, for $\alpha = 0.8$, it has collapsed completely, filling up the initial hole, thus giving rise to a homogeneous profile along the inner part of the microgel. To be more quantitative, we perform simulations at many different values of $\alpha$, finding that the hole fills at $\alpha_{fill} \approx 0.71$. This value is higher than the temperature of the VPT at

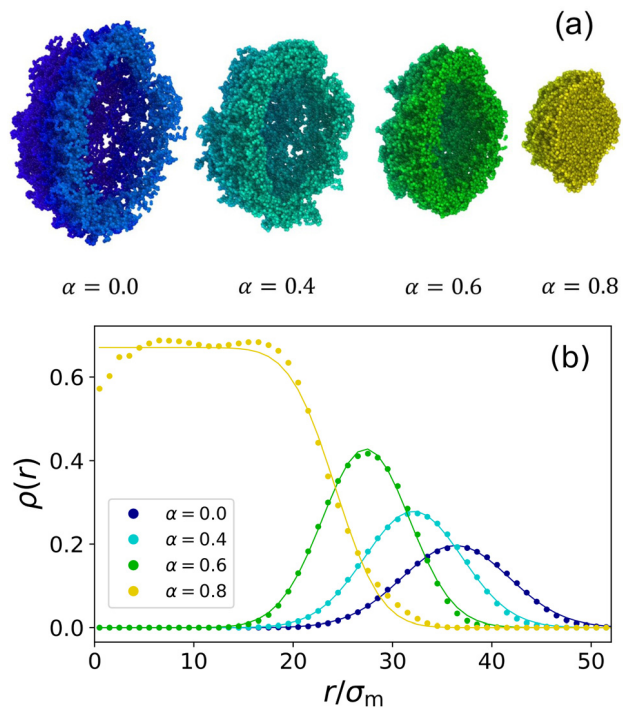

**Fig. 3** (a) Snapshots of a hollow microgel with crosslinker percentage $c = 5\%$ and $\delta_{rel} = 0.275$ obtained from the assembly with $Z_{out} = 60\sigma_m$ and $Z_{in} = 43.5\sigma_m$ ($N_m \approx 44\,000$), at different effective temperatures: $\alpha = 0$, 0.4, 0.6, 0.8 from left to right. A half-slice of the microgels is shown in order to better visualise the presence of the cavity. (b) Radial density profiles $\rho(r)$ at the same temperatures with the corresponding colour coding used consistently throughout this work. Together with the numerical data (symbols), fits (lines) are also reported using a Gaussian function (eqn (11)) in the presence of the cavity and using the fuzzy sphere model when this is filled up ($\alpha = 0.8$).

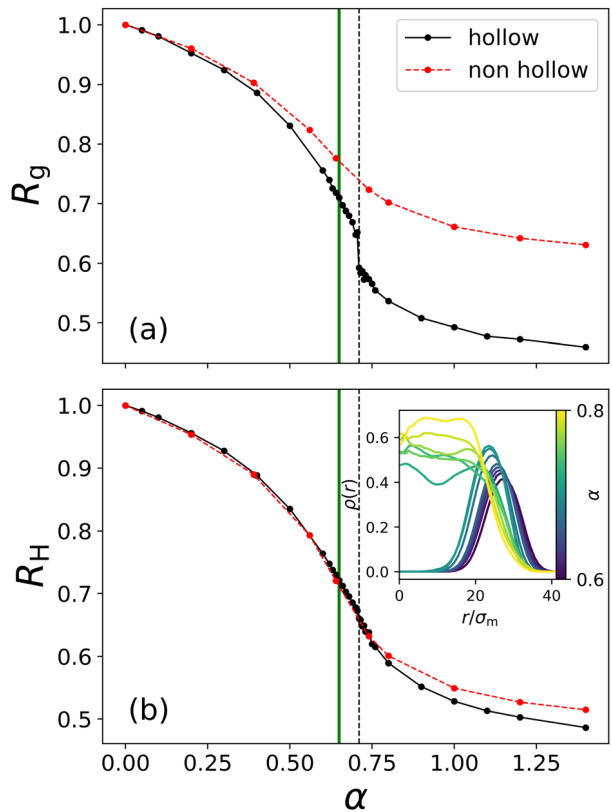

**Fig. 4** Swelling curves of the hollow microgel from Fig. 3 as well as of a non-hollow microgel of comparable size: (a) radius of gyration and (b) hydrodynamic radius *versus* effective temperature $\alpha$. Lines between symbols are guides to the eye. The vertical full line (green) marks the approximate location of the VPT $\alpha_{VPT}$, while the vertical dashed line (black) represents the point of cavity loss of the hollow microgel $\alpha_{fill}$. The inset plot in (b) shows many different density profiles of the hollow microgel at different temperatures ($\alpha = 0.6$, 0.62, 0.63, 0.64, 0.65, 0.68, 0.7, 0.705, 0.71, 0.715, 0.72, 0.73, 0.75, 0.8). At $\alpha = 0.705$, we still measure a clear hollow structure and at $\alpha = 0.71$ the microgel's monomers fill the cavity, and the microgel changes shape.

$\alpha_{VPT} \approx 0.65$. These two quantities shall not be confused: $\alpha_{VPT}$ marks the inflection point of the swelling curve $R_H(\alpha)$ which is the same for different microgels looking at Fig. 4(b) since it depends on the intrinsic change that the polymer chains in the microgel experience due to the temperature-induced coil-globule-transition, whereas $\alpha_{fill}$ only occurs in hollow microgels and varies depending on the microgel's parameters because it refers to the temperature where the microgel's overall shell elasticity is no longer able to sustain the decreasing solvability and crumbles. We thus separately monitor the behaviour of the radius of gyration and of the hydrodynamic radius as a function of effective temperature, both reported in Fig. 4, where one can observe a peculiar feature of hollow microgels. Indeed, $R_g$ is found to display a discontinuity at $\alpha \approx \alpha_{fill}$ since at this point the structure completely changes its topology. This is reflected in the discontinuous behaviour of the density profiles, reported for values of $\alpha$ from just below to just above $\alpha_{fill}$ in the inset of Fig. 4(b). Of course, the non-hollow microgel does not exhibit such a feature, showing a much less pronounced swelling. Indeed, for larger values of $\alpha$, the two curves remain significantly distinct between hollow and non-hollow microgels, with the latter being consistently larger than the hollow case. On the other hand, focusing on the evolution of $R_H$, we find a rather similar swelling curve for both types of microgels at all temperatures and the absence of a discontinuity for the hollow case. Here, the filling of the cavity is smoothly absorbed in the variation of the total size. This difference in the behaviour of $R_g$ and $R_H$ is due to the fact that we calculate the radius of gyration using the coordinates of all monomers in the microgel, while for the hydrodynamic radius we only consider the coordinates of its convex hull which does not include monomers near the hole. Thus, in the former case, the discontinuity is caused by the cavity loss, while in the latter one, the microgel's outer border shrinks continuously with temperature increase, independently of what happens to the hole. These results are in agreement with experimental dynamical light scattering (DLS) measurements for hollow microgels which have reported a continuous variation for $R_H$ as a function of temperature.[17,19,24] To date, experimental measurements of the variation of $R_g$, for example by static light scattering, have not yet been reported to our knowledge for hollow microgels.

## 3.2 How hollow is a hollow microgel?

Having established the main characteristic features of a representative case study of a hollow microgel, we now try to answer numerically this very important question, already asked experimentally some years ago by Dubbert and coworkers.[17] To this end, we vary several parameters in the assembly of the microgels to adjust the balance between shell stiffness and thermoresponsiveness. Namely, we perform a large number of simulations changing the shell thickness, the overall microgel size, and the crosslinker concentration. Size and shell thickness can be modified in experiments by using different sizes for the initial sacrificial cores and coating them with different amounts of monomers, the crosslinker concentration is the standard control parameter to tune the microgel's intrinsic softness.[17,24,26]

We thus initially fix the assembly borders and end up with a rather small relative shell thickness, $\delta_{rel} = 0.21$, and vary the fraction of crosslinkers from 1% to 10%. The corresponding density profiles are reported in Fig. 5 from $\alpha = 0$ to $\alpha = 0.6$ for three values of the crosslinker concentration. We find that for $c = 1\%$ the shell is not able to sustain itself, resulting in a structure that contains monomers within the cavity already at $\alpha = 0.0$. Instead, microgels with $c = 5\%$ and $c = 10\%$ can retain their cavity, at least up to the VPT, even with such a thin shell. As expected, the microgel with a higher crosslinker percentage is stiffer. It is now interesting to ask whether the filling of the cavity still occurs, or if the enhanced stiffness can retard it to the collapsed region or even to completely avoid it. This is a very important question from an application standpoint, as the microgels could preserve their hollowness throughout the volume phase transition. Hence, for the two stiffer microgels of Fig. 5, we report the radius of gyration as a function of temperature in Fig. 6(a). Only for $c = 10\%$, we detect a clear discontinuity in correspondence with the cavity filling. Instead, such a discontinuity is only barely visible for $c = 5\%$, not because the cavity remains intact, but because this happens more or less in correspondence with the VPT, and the simultaneous thermal deswelling does not allow for distinction of the filling event. So, it appears that for small shell thickness, even with enhanced stiffness implemented *via* the presence of the crosslinkers, thermoresponsiveness wins over elasticity, thereby closing the hole and making the microgel non-hollow for temperatures close to or just above the VPT temperature.

We then turn to analyse the opposite case of a much larger relative thickness, $\delta_{rel} = 0.50$, for the same values of crosslinker concentrations and $Z_{out}$. The corresponding swelling curves are reported in Fig. 6(b) and (c). Interestingly, for $c = 1\%$ plotted in Fig. 6(b), the microgel is still able to maintain a cavity up to the VPT. However, the softness of the microgel is such that the filling takes place smoothly, as shown in the inset, and there is no sign of a discontinuity in the swelling curve. Instead, for the more crosslinked microgels, reported in (c), the cavity filling happens at significantly higher values of $\alpha$, so that no discontinuity in $R_g$ is detected and the relative swelling is much more reduced compared to the case of a smaller thickness, even after filling has occurred.

These findings demonstrate that shell thickness is a control parameter that can be used alternatively to crosslinker concentration to tune the softness of the hollow microgels. Indeed, increasing $\delta_{rel}$ keeping the same $c$ has a similar effect as increasing $c$ at fixed $\delta_{rel}$ on the swelling process by moving $\alpha_{fill}$ to higher values. Our findings are in qualitative agreement with available experimental works. In particular, Contreras-Cáceres *et al.* observed that an increased stiffness, realised through a higher crosslinker percentage, reduces the microgel's thermoresponsiveness.[24] The same has been discovered by Dubbert *et al.*,[17] who additionally found that an increase in shell thickness also causes the microgel to become less sensitive to temperature.

Our systematic study, however, introduces new knowledge in the process of cavity filling by tuning its occurrence with respect to the temperature location of the VPT. Indeed, it is important to note that the VPT for PNIPAM microgels always happens at the same temperature $\sim 32$ °C independently of the microgel properties, since it is an underlying property of PNIPAM and of its affinity to water. Similarly, in our simulations, the VPT always occurs at the same effective temperature $\alpha_{VPT} \approx 0.65$, as documented in previous works,[29,35,44] being an intrinsic property of the used solvophobic potential.[36] Therefore, we can decouple the location of $\alpha_{fill}$, which depends on the hollow microgel properties, from that of $\alpha_{VPT}$ which is constant. The three panels in Fig. 6 thus refer to different situations of their relative locations and represent different behaviours. In particular, if $\alpha_{fill}$ lies close enough to the VPT value (panel a) it will be possible to observe a clear discontinuity in $R_g(\alpha)$ (as also observed in Fig. 4(a)), otherwise there will be no visible jump in the swelling curve and a purely continuous behaviour of $R_g$ (Fig. 6(b)). However, if the cavity loss happens for temperatures well above the VPT, as in Fig. 6(c), $R_g(\alpha)$ is found to display a small bump related to $\alpha_{fill}$ so that the filling is still detectable. This feature is highlighted in the inset of Fig. 6(c). The phenomenology is quite interesting and can be related to the findings of a buckling transition of hollow microgels upon increasing osmotic pressure from non-adsorbed polymer chains.[21] Here, of course, we are examining a microgel on its own. Still, it is legitimate to ask whether the competition between the elasticity of the shell and thermoresponsiveness could play a role in modifying the microgel structure. We thus examine the most rigid case in more detail close to the occurrence of the bump, where $R_g$ actually increases again with $\alpha$, signaling some kind of modification. The corresponding snapshot of the microgel with $c = 10\%$ at $\alpha = 1.2$ is also shown in Fig. 6(c), displaying a microgel that has completely lost its sphericity, adopting a quasi-2D oblate shape. This causes the radius of gyration to slightly increase, before the microgel shrinks further at even higher temperatures, creating a small bump in the swelling curve of thick and highly crosslinked shells. A microgel with the same $c$ but a much smaller $\delta_{rel}$, reported in the snapshots of Fig. 6(a), instead takes on a bowl-like shape just above the filling, appearing very close to a buckling transition before it eventually collapses into a small sphere for higher values of $\alpha$. It will thus be interesting in the future to deepen the investigation of these microgels under crowded conditions, either by adding polymer chains as in ref. 21 or by increasing microgel concentration.

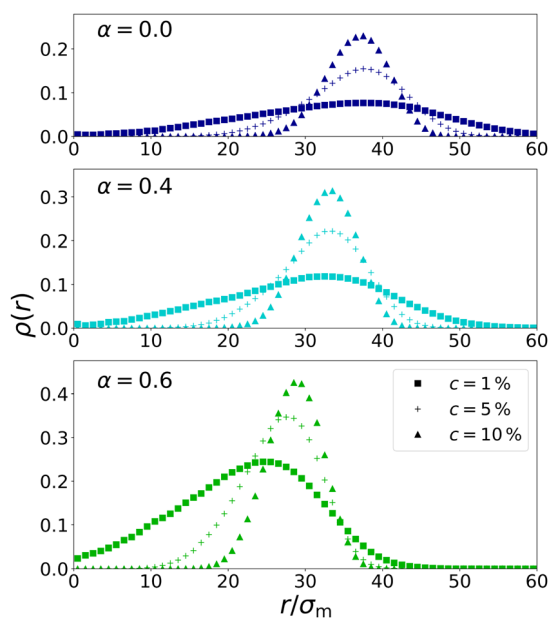

**Fig. 5** Radial density distributions for microgels with different crosslinker concentrations $c$ at different temperatures (see labels). The considered microgels have the same relative shell thickness $\delta_{rel} = 0.21$ and number of monomers $N_m \approx 41\,000$ (initial boundaries: $Z_{in} = 50\sigma_m$ and $Z_{out} = 63\sigma_m$).

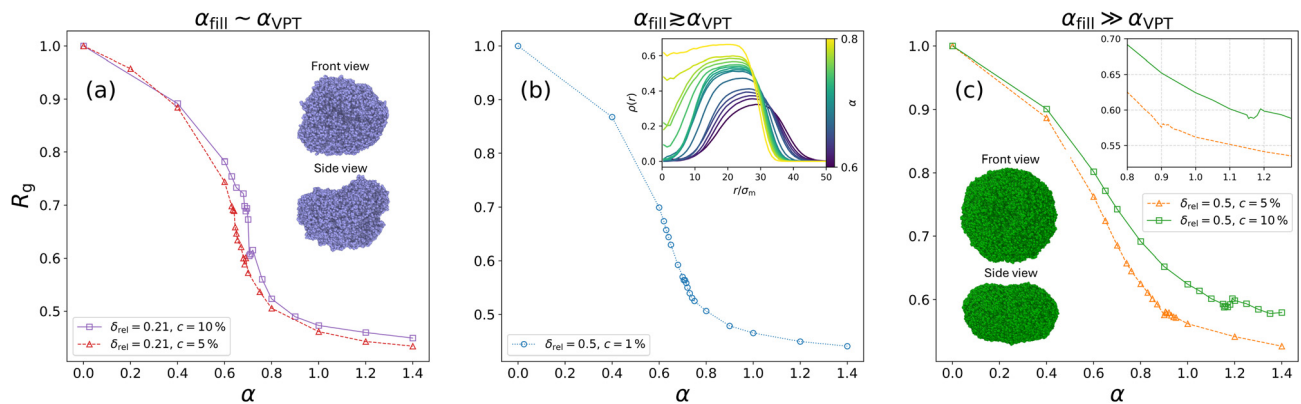

**Fig. 6** Swelling curves of different microgels in terms of $R_g$. All the microgels have the same initial outer assembly border $Z_{out} = 63\sigma_m$ while they differ in shell thickness $\delta_{rel}$ and crosslinker concentration $c$; (a) $\delta_{rel} = 0.21$ and $c = 5$, 10% microgels—thin and elastic shells—for which $\alpha_{fill} = 0.645$ for $c = 5\%$, exhibiting a slight discontinuity in $R_g$ and $\alpha_{fill} = 0.705$ for $c = 10\%$, which displays a clear jump. The snapshots refer to the $c = 10\%$ case just above the cavity loss ($\alpha = 0.8$); (b) $\delta_{rel} = 0.5$ and $c = 1\%$ microgel which fills its cavity at $\alpha_{fill} = 0.72$ in a continuous way, emphasised by the inset showing the different $\rho(r)$ with varying $\alpha$; (c) $\delta_{rel} = 0.5$ and $c = 5$, 10% microgels—thick and highly crosslinked shells—with $\alpha_{fill} = 0.91$ for $c = 5\%$ and $\alpha_{fill} = 1.155$ for $c = 10\%$, where a small bump is detected in $R_g$ – highlighted in the inset and discussed in the text. The snapshots refer to $c = 10\%$ microgel at $\alpha = 1.2$, just above its $\alpha_{fill}$.

### 3.3 Dependence on microgel size

Since our simulations are based on coarse-grained monomer modeling, which of course implies the use of much smaller microgels than in experiments, we now try to assess how $\alpha_{fill}$ is affected by increasing the number of monomers $N_m$ in the microgel coarse-graining. Fig. 7(a) thus reports $\alpha_{fill}$ versus size for many different microgels. Since $N_m$ can be the same for two microgels with different $\delta_{rel}$ and $c$, the size is expressed in terms of $Z_{out}$, while we report in Table 1 the parameters of all studied microgels, explicitly including $N_m$.

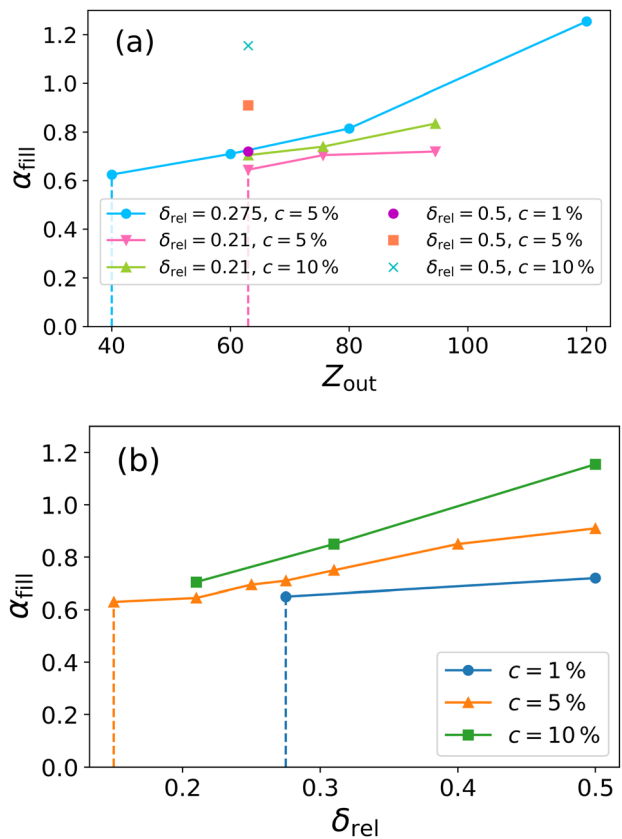

**Fig. 7** Cavity loss point $\alpha_{fill}$ for various hollow microgels of different sizes, shown as a function of initial outer radius $Z_{out}$ in (a) and of shell thickness $\delta_{rel}$ in (b). Symbols refer to measured data points; solid lines connect microgels with the same crosslinker concentration $c$. All $\alpha_{fill}$ points are estimated with an accuracy of 0.005 in $\alpha$. The dashed vertical lines mark the approximate limit for the corresponding parameter ($Z_{out}$ or $\delta_{rel}$) below which it becomes impossible, using the same colour-coded crosslinker percentage, to simulate a hollow microgel that does not show monomers inside the cavity at $\alpha = 0$.

**Table 1** Table of input shell border parameters for different $\delta_{rel}$ cases of the studied microgels shown in Fig. 7(a), including their monomer number. Lengths are given in units of $\sigma_m$

| $\delta_{rel}$ | Values of $Z_{in}$, $Z_{out}$ and $N_m$ |
|---|---|
| 0.5 | $Z_{in} = 31.5$, $Z_{out} = 63 \rightarrow N_m \sim 73\,000$ |
| 0.275 | $Z_{in} = 29$, $Z_{out} = 40 \rightarrow N_m \sim 13\,000$ |
| | $Z_{in} = 43.5$, $Z_{out} = 60 \rightarrow N_m \sim 44\,000$ |
| | $Z_{in} = 58$, $Z_{out} = 80 \rightarrow N_m \sim 105\,000$ |
| | $Z_{in} = 87$, $Z_{out} = 120 \rightarrow N_m \sim 355\,000$ |
| 0.21 | $Z_{in} = 50$, $Z_{out} = 63 \rightarrow N_m \sim 41\,000$ |
| | $Z_{in} = 60$, $Z_{out} = 75.6 \rightarrow N_m \sim 72\,000$ |
| | $Z_{in} = 75$, $Z_{out} = 94.5 \rightarrow N_m \sim 140\,000$ |

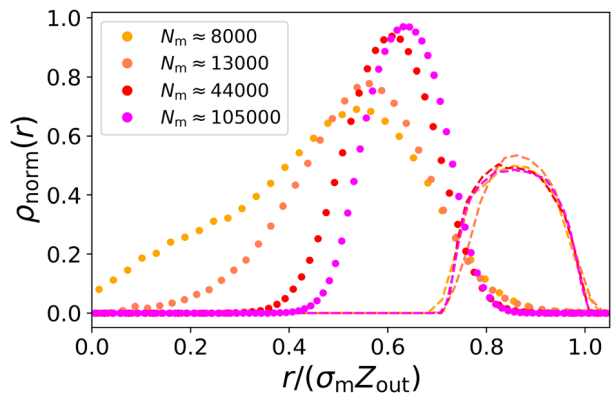

**Fig. 8** Normalized density profiles of four different microgels with the same $\delta_{rel} = 0.275$ and $c = 5\%$ and different size $N_m$. From left to right, they correspond to: $Z_{out} = 34.5\sigma_m$, $40\sigma_m$, $60\sigma_m$, $80\sigma_m$. The dotted lines are the equilibrated profiles at $\alpha = 0$, while the dashed lines represent the initial assembly configuration, resulting from the patchy particle simulation done with oxDNA. To better visualise the shift with size, data have been $x$-scaled to the corresponding $Z_{out}$ and $y$-scaled to a unit integral.

Fig. 7(a) clearly shows that, by increasing $Z_{out}$ and hence $N_m$, the cavity fills in at higher and higher values of effective temperature. There does not seem to be a saturation microgel size, particularly for the reference case $\delta_{rel} = 0.275$ and $c = 5\%$, that is studied up to very large $N_m$. This result is in agreement with experimental studies of very large hollow microgels, which are found to maintain the cavity open even well above the VPT[19] or even to almost lose thermoresponsiveness completely.[21]

To better visualise the dependence of cavity filling on the shell thickness, we also report $\alpha_{fill}$ as a function of $\delta_{rel}$ in Fig. 7(b), showing results for different crosslinker concentrations. An increase in shell thickness, thus in microgel elasticity, has the clear effect of stabilising the hole up to very high temperatures. The opposite is true upon decreasing crosslinker concentration. Importantly, we find that, for high enough $\delta_{rel}$, also $c = 1\%$ microgels can be stabilized as hollow ones up to temperatures around the VPT, something not yet explored in experiments. These results suggest that an accurate balance between $\delta_{rel}$ and $c$ allows one to tune the hollowness of the microgel to the desired temperature, according to the needs of applications.

Having established that the use of a large number of monomers helps to stabilise the cavity, it is now important to ask the opposite question, namely, how small a microgel can be while maintaining its cavity. Indeed, the stability of the cavity is an important challenge in establishing a minimal number of monomers to be used in computationally feasible bulk simulations. For simplicity, we focus on our reference case $\delta_{rel} = 0.275$ and $c = 5\%$ and vary $N_m$ in order to establish the lowest number of monomers necessary to observe a clear, stable cavity.

Fig. 8 shows the normalized density profiles $\rho_{norm}(r)$ equilibrated at $\alpha = 0$ (dots) of representative microgels with $N_m$ ranging from $\sim 8000$ to $\sim 105\,000$. It is immediately evident that the smallest microgel is not hollow, while the cavity becomes more and more pronounced as $N_m$ increases. This can be best appreciated by comparing the density profiles, where the microgel is fully equilibrated in good solvent, with the initial configurations resulting from the assembly (dashed lines). The data have all been scaled to relative distances in order to highlight how the change in shape caused by equilibration depends on microgel size. Hence, for very large $N_m$, we observe a slight shift of the network distribution closer to the microgel's center of mass, while still preserving a symmetric, almost Gaussian, density profile with a relative thickness very similar to the nominal one from the assembly. As $N_m$ decreases, $\rho(r)$ is found to shift more and more to the center away from the (fixed) initial density profile until it becomes slightly asymmetric for $N_m \sim 13\,000$ and finally loses the cavity completely below this threshold number of monomers. Incidentally, despite seeming just barely able to sustain the cavity, checking the profile for $N_m \sim 13\,000$ at different $\alpha$ (not shown), the microgel has just enough elasticity to keep the hollowness up to the VPT. This is due to the fact that, in general, the microgel gains additional stiffness by increasing temperature, which helps to stabilize the cavity. Therefore, the present investigation suggests that for the chosen values of $\delta_{rel} = 0.275$ and $c = 5\%$, we need at least $N_m \sim 13\,000$ to be able to simulate truly hollow microgels. This number is of the order of $10^4$ monomers, which is a low enough number to allow for multiple microgel simulations within present computational resources.[12]

### 3.4 Form factor analysis and comparison to experiments

In this last section, we focus on reporting the form factors $P(q)$, directly calculated from the simulations, for different hollow microgels. We describe them with the models explained in the Methods section. We then try to compare some numerical predictions with available experimental data.

We show $P(q)$ for hollow microgels with $c = 5\%$ and two different relative thicknesses in Fig. 9. In particular, data refer to relatively thin microgels ($\delta_{rel} = 0.275$) in panel (a) and to thicker shells in panel (b). We use two approaches to describe the data: (i) we fit the corresponding density profiles, then evaluate the Fourier transform with the obtained fit parameters, and (ii) we directly fit the form factors and compare the resulting fit parameters to the previous ones. The two approaches turn out to be indistinguishable from each other, confirming the robustness of the results. We find that the simple Gaussian form factors capture the behaviour of the symmetric thinner microgels reported in eqn (12) quite well, while the parabolic model with $R_{in} = R_{out} - s_{out} - s_{in}$, i.e., $W = 0$, as explained in the Methods, describes them with equal precision while $s_{in} \approx s_{out}$. This is true as long as the cavity stays open, for $\alpha \lesssim \alpha_{fill}$, but for higher temperatures it is necessary to use the fuzzy sphere model. The change in topology for $P(q)$ is clearly evident in Fig. 9(a) for $\alpha = 0.8$ compared to the other temperatures since it displays less sharp minima that indicate a fuzzy-sphere-like structure. Next, we examine thicker hollow microgels in Fig. 9(b), where the Gaussian shape of the density profiles no longer holds. To describe the form factor of the thicker microgel with $\delta_{rel} = 0.5$, we then resort to the parabolic approach with $R_{in} = R_{out} - s_{out} - s_{in} - W$ for both the density profiles and the form factors (eqn (8) and (10)). This procedure is now applied to all $\alpha$ reported, because this microgel maintains its hollow character even at $\alpha = 0.8$.

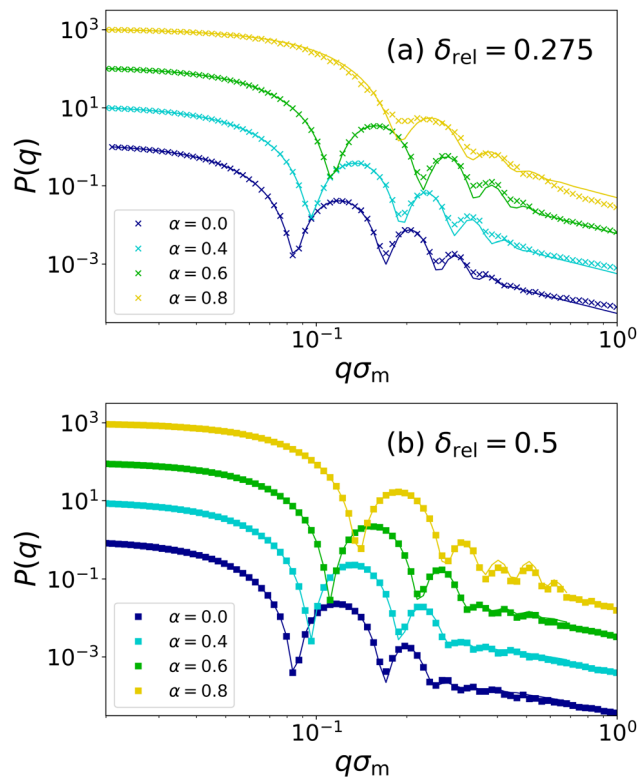

**Fig. 9** Form factors $P(q)$ for microgels with $c = 5\%$ and (a) $\delta_{rel} = 0.275$ and (b) $\delta_{rel} = 0.5$ for different values of $\alpha$, arbitrarily shifted vertically to improve visualisation. Symbols are numerical data and lines are fits obtained using the Gaussian approximation in eqn (12) for the curves in (a) with $\alpha = 0$–0.6, the fuzzy sphere model for $\alpha = 0.8$, and finally the parabolic approach in (b) for all $\alpha$.

We summarize all the fit parameters for the two sets of microgels at the different values of $\alpha$, illustrated in Fig. 9, in Table 2. Parameters are listed both resulting from density profiles and from form factor fits, confirming that the two methods yield very similar results. Similarly, a consistency is obtained between Gaussian and parabolic (with $R_{in} = R_{out} - s_{out} - s_{in} \to W = 0$) functional forms for the thinner microgels. Of course, for small enough microgels, the Gaussian approach can be preferred, especially for describing numerical data, due to the smaller number of fit parameters involved.

We can now directly compare the simulation results to experimental findings. We examine the experiments of Hazra and coworkers[19] since these hollow microgels have a shell thickness and crosslinker percentage close to our reference case ($\delta_{rel} \approx 0.275$ and $c \approx 5\%$). The experimental form factors of these rather large microgels, measuring approximately 1 µm in the swollen state, have been measured by static light scattering at two different temperatures, 20 °C and 40 °C. The comparison between experimental and numerical form factors at these two temperatures is reported in Fig. 10, where the numerical curves have been shifted on $q$-axis by a common factor for each temperature, identifying the value of $\sigma_m$ that is used in the coarse-graining approach (see caption of Fig. 10). After rescaling the position of the first peak, we find that the subsequent peaks also match the positions of the experimental form factors, especially for the largest simulated microgel ($N_m \approx 355\,000$), allowing us to identify that $\alpha = 0.0$ corresponds to $T = 20$ °C, while $\alpha = 0.65$ is consistent with $T = 40$ °C. The comparison confirms that the numerical model is able to well describe the experimental behaviour and, most importantly, to reproduce the presence of the cavity for both temperatures.[19]

**Table 2** Full set of parameters obtained from fitting $\rho(r)$ (with an added index $\rho$) or $P(q)$ (same symbols without index) using a parabolic fit for $W > 0$ for the density profile (eqn (8) with $R_{in} = R_{out} - s_{out} - s_{in} - W$) of the microgel with $\delta_{rel} = 0.50$ ($Z_{in} = 31.5\sigma_m$ and $Z_{out} = 63\sigma_m$), while using both a Gaussian (eqn (11)) and a parabolic fit with $W = 0$ (eqn (8) with $R_{in} = R_{out} - s_{out} - s_{in}$) for $\delta_{rel} = 0.275$ ($Z_{in} = 43.5\sigma_m$ and $Z_{out} = 60\sigma_m$) up to $\alpha = 0.6$. Lengths are given in units of $\sigma_m$

| $\delta_{rel} = 0.50$, $N_m \approx 73\,000$ ($Z_{in} = 31.5\sigma_m$ and $Z_{out} = 63\sigma_m$ (eqn (8) with $R_{in} = R_{out} - s_{out} - s_{in} - W$)) | | | | | | | | | |
|---|---|---|---|---|---|---|---|---|---|
| $\alpha$ | $s_{in,\rho}$ | $s_{in}$ | $s_{out,\rho}$ | $s_{out}$ | $R_{out,\rho}$ | $R_{out}$ | $W_\rho$ | $W$ | $a_\rho$ | $a$ |
| 0.0 | 8.27 ± 0.07 | 8.10 ± 5.28 | 7.37 ± 0.07 | 9.40 ± 2.24 | 46.58 ± 0.02 | 46.07 ± 1.09 | 6.45 ± 0.12 | 3.70 ± 9.62 | 0.20 ± 0.00 | 0.23 ± 0.03 |
| 0.4 | 7.49 ± 0.06 | 8.60 ± 4.43 | 6.61 ± 0.06 | 8.57 ± 4.43 | 41.33 ± 0.02 | 40.73 ± 1.11 | 5.53 ± 0.11 | 1.29 ± 8.74 | 0.29 ± 0.00 | 0.33 ± 0.04 |
| 0.6 | 6.35 ± 0.04 | 6.78 ± 3.54 | 5.72 ± 0.04 | 6.38 ± 1.52 | 35.71 ± 0.01 | 35.59 ± 0.65 | 5.26 ± 0.07 | 3.89 ± 6.32 | 0.44 ± 0.00 | 0.47 ± 0.04 |
| 0.8 | 4.78 ± 0.02 | 4.15 ± 5.33 | 4.19 ± 0.02 | 3.41 ± 1.52 | 29.96 ± 0.02 | 30.15 ± 0.35 | 8.44 ± 0.03 | 10.21 ± 7.23 | 0.69 ± 0.00 | 0.71 ± 0.03 |

| $\delta_{rel} = 0.275$, $N_m \approx 44\,000$, ($Z_{in} = 43.5\sigma_m$ and $Z_{out} = 60\sigma_m$ (eqn (8) with $R_{in} = R_{out} - s_{out} - s_{in}$ and eqn (11))) | | | | | | | | |
|---|---|---|---|---|---|---|---|---|
| $\alpha$ | $s_{in,\rho}$ | $s_{in}$ | $s_{out,\rho}$ | $s_{out}$ | $R_{out,\rho}$ | $R_{out}$ | $a_\rho$ | $a$ |
| 0.0 | 7.21 ± 0.08 | 5.94 ± 1.76 | 6.37 ± 0.07 | 8.23 ± 1.33 | 43.25 ± 0.03 | 43.25 ± 0.43 | 0.29 ± 0.00 | 0.21 ± 0.01 |
| 0.4 | 6.51 ± 0.07 | 6.12 ± 1.37 | 5.97 ± 0.07 | 6.91 ± 1.07 | 38.32 ± 0.03 | 38.52 ± 0.31 | 0.27 ± 0.00 | 0.28 ± 0.01 |
| 0.6 | 5.82 ± 0.05 | 5.24 ± 1.13 | 5.41 ± 0.05 | 6.47 ± 0.85 | 32.92 ± 0.02 | 33.02 ± 0.27 | 0.41 ± 0.00 | 0.42 ± 0.01 |

| $\alpha$ | $\sigma_\rho$ | $\sigma$ | $\mu_\rho$ | $\mu$ | $C_\rho$ | $C$ |
|---|---|---|---|---|---|---|
| 0.0 | 5.40 ± 0.02 | 5.14 ± 0.08 | 36.35 ± 0.02 | 36.31 ± 0.04 | 0.20 ± 0.00 | 0.23 ± 0.00 |
| 0.4 | 4.87 ± 0.01 | 4.66 ± 0.06 | 32.10 ± 0.01 | 32.05 ± 0.03 | 0.28 ± 0.00 | 0.31 ± 0.00 |
| 0.6 | 4.38 ± 0.01 | 4.22 ± 0.05 | 27.31 ± 0.02 | 27.24 ± 0.03 | 0.43 ± 0.00 | 0.46 ± 0.00 |

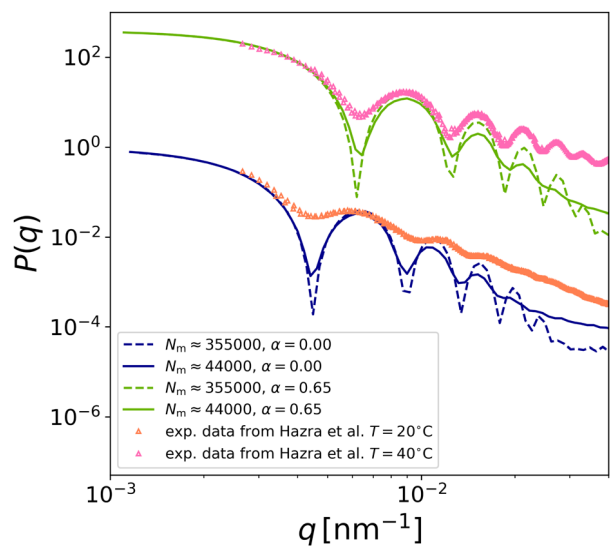

**Fig. 10** Comparison of numerical (lines) and experimental form factors (symbols). The latter are taken from ref. 19 for hollow microgels with $\delta_{rel} \sim 0.275$ and $c \sim 5\%$. The simulation curves have been rescaled not only on the $y$-axis but also on the $x$-axis using the factor $q_{exp} = q_{sim}/\sigma_m$ with $\sigma_m \approx 4$ nm for the smaller microgel ($Z_{in} = 60\sigma_m$ and $Z_{out} = 43.5\sigma_m$) and $\sigma_m \approx 8.9$ nm for the biggest one studied here: $Z_{in} = 87\sigma_m$ and $Z_{out} = 120\sigma_m$. Data have been shifted vertically to improve visualisation.

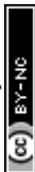

## 4 Discussion and conclusions

In this work, we took advantage of computer simulations of a realistic microgel model to investigate in detail the swelling behaviour of hollow microgels with different characteristics across the volume phase transition. In an attempt to quantify the effects of softness and thermoresponsiveness on the hollowness of the microgel and on how this can persist even beyond the VPT, we varied the main control parameters that are used in experiments. In particular, we tuned the crosslinker concentration and the relative shell thickness to modulate the filling of the cavity in relation to the occurrence of the VPT. While the latter is fixed by the polymer's affinity to the solvent to always occur at the same temperature, the former can thus be experimentally adjusted. Our simulations found that the filling manifests itself, under appropriate conditions, as a discontinuous behaviour of the radius of gyration as a function of temperature. In particular, we found that increasing the crosslinker concentration has qualitatively a similar effect to increasing the shell thickness. However, while $c$ acts at a local level, defining the overall stiffness of the shell, $\delta_{rel}$ dictates the topology of the microgel. Thus, a larger shell thickness provides an increase in the value of $W$ in Fig. 2, yielding a microgel with a different structure than for thin shells, where $W = 0$, indicating a transition from parabolic to Gaussian density profile. This qualitative change also impacts the swelling process, making a sharp discontinuity in $R_g(\alpha)$ much more likely for thinner shells (see Fig. 6(a)). Therefore, such a discontinuity should be looked for in experiments by focusing on microgels with thin shells and a high enough crosslinker percentage, allowing them to maintain the cavity just above the VPTT. Importantly, we do not observe a similar feature in the swelling behaviour of the hydrodynamic radius, due to the stronger influence of the outer chains on this quantity and possibly also to the reduced numerical system size, which may disfavour the observation of a discontinuity in the total microgel size. Indeed, experimental measurements of $R_H$ have not detected such a behaviour so far, but it may be worth searching for it by directly measuring $R_g$.

In order to assess the importance of size effects in our study, we prepared microgels of different sizes and numbers of monomers $N_m$ ranging between $10^4$ and $10^5$. We then reported several of the microgel properties as a function of $N_m$, finding that increasing the size enhances the stiffness of the shell with respect to cavity filling. However, we also determined $\sim 10^4$ as the lower limit of stability of hollow microgels for realistic numerical modelling in future studies. Indeed, once prepared with a nominal cavity, all initial confinements are removed, as in experiments by the dissolution of the core, the shell tends to partially fill the hole, as also seen experimentally by Dubbert *et al.*[17] who monitored the structure of the microgels before and after core dissolution. A nice correspondence between experimental and numerical form factors has been found comparing our results to the hollow microgels, recently synthesised by Hazra and coworkers.[19] In addition, the dependence of the cavity filling temperature $\alpha_{fill}$ on the crosslinker concentration or shell thickness was found to be in qualitative agreement with available experiments.[17,24] These results give us confidence in the numerical model, opening the way to extend our analysis to the collective behaviour of hollow microgels under crowded conditions. A first study in this direction was put forward in ref. 45 where the behaviour of hollow microgels was studied in a binary mixture of hollow and regular microgels, revealing through simulations that for the studied parameters hollow microgels experience an effective potential that is softer than for regular ones. Another interesting direction to pursue in the future is to assess the role of charges on the swelling response of hollow microgels, as initially done by Hazra and coworkers.[19] However, it is already known that the presence of ionic groups favours the stretching of the network[40] and thus should effectively act as an additional contribution to the stiffness of the network. Their presence could thus modify the response to external stimuli such as pH, which could hold promises for practical purposes.

Finally, the present study suggests that different behaviours are observed when the elasticity of microgels is enhanced. For example, the microgels have a stronger tendency to deform, as in the case of $c = 10\%$ (see snapshots of Fig. 7). Such interesting properties may thus be enhanced by the presence of charged groups within the shell. We also explored the very soft case of hollow microgels with $c = 1\%$, which was found to be able to sustain its cavity up to the VPT for a large enough shell thickness. This avenue could be further explored in experiments, where mostly microgels with $c \geq 5\%$ have been considered so far. In summary, our work can provide useful guidelines for both experiments and simulations in selecting the best parameters in the synthesis of the microgels to enhance or tune their hollowness at a desired temperature and to optimize their use for specific applications.

## Author contributions

Author contributions are defined based on CRediT (Contributor Roles Taxonomy). Conceptualization: E. Z.; formal analysis: L. R., E. Z.; funding acquisition: E. Z.; investigation: L. R., E. Z.; methodology: L. R., E. Z.; project administration: E. Z.; supervision: E. Z.; validation: L. R., E. Z.; visualisation: L. R.; writing – original draft: L. R., E. Z.; writing – review and editing: L. R., E. Z.

## Data availability

Data for this article are available at Zenodo at **https://doi.org/10.5281/zenodo.15125274**.

## Conflicts of interest

There are no conflicts to declare.

## Appendices

### Fourier transformation of parabolic density distribution

Since for the parabolic approximation of the density profile, $\rho_{\text{parab}}(r)$ is made up of $i = 4$ consecutive parabolas (*cf.* eqn (8)) which we call here $\rho_i(r)$, the full amplitude $A_{\text{parab}}(q)$ contains four terms:

$$
\begin{aligned}
A_1(q) &= 4\pi \int_{R_{\text{in}}-s_{\text{in}}}^{R_{\text{in}}} \rho_1(r) \frac{\sin qr}{qr} r^2 \mathrm{d}r \\
&= \frac{1}{2q^5 s_{\text{in}}^2} a\big(\big(-q^3 R_{\text{in}} s_{\text{in}}^2 + 2q(R_{\text{in}} + 2s_{\text{in}})\big) \cos(qR_{\text{in}}) \\
&\quad - 2q(R_{\text{in}} - s_{\text{in}}) \cos(q(R_{\text{in}} - s_{\text{in}})) - 6\sin(qR_{\text{in}}) \\
&\quad + 2q^2 R_{\text{in}} s_{\text{in}} \sin(qR_{\text{in}}) + q^2 s_{\text{in}}^2 \sin(qR_{\text{in}}) \\
&\quad + 6\sin(q(R_{\text{in}} - s_{\text{in}}))\big),
\end{aligned}
\tag{13}
$$

$$
\begin{aligned}
A_2(q) &= 4\pi \int_{R_{\text{in}}}^{R_{\text{in}}+s_{\text{in}}} \rho_2(r) \frac{\sin qr}{qr} r^2 \mathrm{d}r \\
&= \frac{1}{2q^5 s_{\text{in}}^2} a\big(q\big(-4s_{\text{in}} + R_{\text{in}}\big(2 + q^2 s_{\text{in}}^2\big)\big) \cos(qR_{\text{in}}) \\
&\quad - 2q(R_{\text{in}} + s_{\text{in}})\big(1 + q^2 s_{\text{in}}^2\big) \cos(q(R_{\text{in}} + s_{\text{in}})) \\
&\quad - \big(6 + q^2 s_{\text{in}}(2R_{\text{in}} + s_{\text{in}})\big) \sin(qR_{\text{in}}) \\
&\quad + 2\big(3 + q^2 s_{\text{in}}^2\big) \sin(q(R_{\text{in}} + s_{\text{in}}))\big),
\end{aligned}
\tag{14}
$$

$$
\begin{aligned}
A_3(q) &= 4\pi \int_{R_{\text{out}}-s_{\text{out}}}^{R_{\text{out}}} \rho_3(r) \frac{\sin qr}{qr} r^2 \mathrm{d}r \\
&= -\frac{1}{2q^5 s_{\text{out}}^2} a\big(q\big(4s_{\text{out}} + R_{\text{out}}\big(2 + q^2 s_{\text{out}}^2\big)\big) \cos(qR_{\text{out}}) \\
&\quad - 2q(R_{\text{out}} - s_{\text{out}})\big(1 + q^2 s_{\text{out}}^2\big) \cos(q(R_{\text{out}} - s_{\text{out}})) \\
&\quad - 6\sin(qR_{\text{out}}) + 2q^2 R_{\text{out}} s_{\text{out}} \sin(qR_{\text{out}}) \\
&\quad - q^2 s_{\text{out}}^2 \sin(qR_{\text{out}}) + 6\sin(q(R_{\text{out}} - s_{\text{out}})) \\
&\quad + 2q^2 s_{\text{out}}^2 \sin(q(R_{\text{out}} - s_{\text{out}}))\big),
\end{aligned}
\tag{15}
$$

$$
\begin{aligned}
A_4(q) &= 4\pi \int_{R_{\text{out}}}^{R_{\text{out}}+s_{\text{out}}} \rho_4(r) \frac{\sin qr}{qr} r^2 \mathrm{d}r \\
&= \frac{1}{2q^5 s_{\text{out}}^2} a\big(q\big(4s_{\text{out}} + R_{\text{out}}\big(-2 + q^2 s_{\text{out}}^2\big)\big) \cos(qR_{\text{out}}) \\
&\quad + 2q(R_{\text{out}} + s_{\text{out}}) \cos(q(R_{\text{out}} + s_{\text{out}})) \\
&\quad + \big(6 + q^2(2R_{\text{out}} - s_{\text{out}})s_{\text{out}}\big) \sin(qR_{\text{out}}) \\
&\quad - 6\sin(q(R_{\text{out}} + s_{\text{out}}))\big),
\end{aligned}
\tag{16}
$$

and in case $W \neq 0$ there is an additional term in between:

$$
\begin{aligned}
A_5(q) &= 4\pi \int_{R_{\text{in}}+s_{\text{in}}}^{R_{\text{out}}-s_{\text{out}}} a \cdot \frac{\sin qr}{qr} r^2 \mathrm{d}r \\
&= \frac{a}{q^3}(q(R_{\text{in}} + s_{\text{in}}) \cos(q(R_{\text{in}} + s_{\text{in}})) \\
&\quad + q(-R_{\text{out}} + s_{\text{out}}) \cos(q(R_{\text{out}} - s_{\text{out}})) \\
&\quad - \sin(q(R_{\text{in}} + s_{\text{in}})) + \sin(q(R_{\text{out}} - s_{\text{out}})))
\end{aligned}
\tag{17}
$$

They can now all be added to make up the total Fourier transformation $A_{\text{parab}}(q) = A_1(q) + A_2(q) + A_3(q) + A_4(q) + (A_5(q))$ which can then be used for eqn (10).

## Acknowledgements

We thank Jerome Crassous and Walter Richtering for providing experimental form factors. We also thank Lorenzo Rovigatti, Rodrigo Rivas-Barbosa, and Yuri Gerelli for useful discussions. We acknowledge financial support from the European Union (HorizonMSCA-Doctoral Networks) through the project QLUS-TER (HORIZON-MSCA-2021-DN-01-GA101072964).

# Notes and references